\documentclass[useAMS,usenatbib]{mn2e}

\usepackage{amssymb}
\usepackage{graphicx}
\newcommand{\chem}[2]{$\mathrm{^{#1}#2}$}
\def\Msun{$M_\odot$}
\def\msun{$M_\odot$}

\def\rsun{$R_\odot$}

\title[Radiation-hydrodynamical modelling of underluminous SNe IIP]{Radiation-hydrodynamical modelling of underluminous type II plateau Supernovae}
\author[M.~L.~Pumo et al.]{M.~L.~Pumo$^{1,2}$\thanks{E-mail: mlpumo@astropa.unipa.it or mlpumo@oact.inaf.it}, L.~Zampieri$^{3}$, S.~Spiro$^{3}$, A.~Pastorello$^{3}$, S.~Benetti$^{3}$, E.~Cappellaro$^{3}$,
\newauthor 
G.~Manic\`o$^{2}$, M.~Turatto$^{3}$\\ 
$^{1}$INAF - Osservatorio Astronomico di Palermo, Piazza del Parlamento 1, 90134 Palermo, Italy\\
$^{2}$Universit\`a degli studi di Catania, Dip.~di Fisica e Astronomia, Via Santa Sofia 78, 95123 Catania, Italy\\
$^{3}$INAF - Osservatorio Astronomico di Padova, Vicolo dell'Osservatorio 5, I-35122 Padova, Italy}
\begin{document}

\date{Accepted 2016 October 8. Received 2016 October 6; in original form 2016 May 16.}

\pagerange{\pageref{firstpage}--\pageref{lastpage}} \pubyear{....}

\maketitle

\label{firstpage}

\begin{abstract}
With the aim of improving our knowledge about the nature of the progenitors of low-luminosity Type II plateau supernovae (LL SNe IIP), we made radiation-hydrodynamical models of the well-sampled LL SNe IIP 2003Z, 2008bk and 2009md. For these three SNe we infer explosion energies of $0.16$-$0.18$ foe, radii at explosion of $1.8$-$3.5 \times 10^{13}$ cm, and ejected masses of $10$-$11.3$\Msun. The estimated progenitor mass on the main sequence is in the range $\sim 13.2$-$15.1$\Msun\, for SN 2003Z and $\sim 11.4$-$12.9$\Msun\, for SNe 2008bk and 2009md, in agreement with estimates from observations of the progenitors.
These results together with those for other LL SNe IIP modelled in the same way, enable us also to conduct a comparative study on this SN sub-group. 
The results suggest that: a) the progenitors of faint SNe IIP are slightly less massive and have less energetic explosions than those of intermediate-luminosity SNe IIP; b) both faint and intermediate-luminosity SNe IIP originate from low-energy explosions of red (or yellow) supergiant stars of low-to-intermediate mass; c) some faint objects may also be explained as electron-capture SNe from massive super-asymptotic giant branch stars; and d) LL SNe IIP form the underluminous tail of the SNe IIP family, where the main parameter ``guiding'' the distribution seems to be the ratio of the total explosion energy to the ejected mass. Further hydrodynamical studies should be performed and compared to a more extended sample of LL SNe IIP before drawing any conclusion on the relevance of fall-back to this class of events.
\end{abstract}                                                                                                                                                                                                                                                                                                                                                                                                                                                                                                                                                                                                                                                                                                                                                                                                                                                                                                                                                                                                                                   


\section{Introduction}
\label{intro}

It is well known that Type II plateau supernovae (SNe IIP hereafter) are explosive events showing hydrogen (and metal) lines with P-Cygni profiles in their spectra, and an extended period (typically the early $\sim$ 100 days of post-explosion evolution) during which the bolometric light curve remains constant to within $\sim$ 0.5 magnitudes \citep*[e.g.][] {filip97,turatto03,turatto07,smartt09,pastorello12}. After this plateau or photospheric phase, the bolometric light curve shows a rapid decay followed by a transition to a linear decline of $\sim$ 0.98 magnitudes every 100 days \citep[e.g.][]{sollerman02}. During this post-explosion period (radioactive-decay phase), the continuum electromagnetic emission is thought to be powered by the energy released from the radioactive decay of \chem{56}{Ni} through the nuclear decay chain \chem{56}{Ni} $\rightarrow$ \chem{56}{Co} $\rightarrow$ \chem{56}{Fe} \citep[e.g.][]{PZ11}.\par

``Typical'' SNe IIP (e.g.~SNe 2004et, 1999em and 1999gi) have P-Cygni line profiles with widths of several thousand km s$^{-1}$ (from $\sim$ 3000 to $\sim$ 15000 km s$^{-1}$) and bolometric luminosities at the plateau from $\sim$ 10$^{42}$ to $\sim$ 10$^{43}$ erg s$^{-1}$ \citep[e.g.][]{sahu06,misra07,maguire10,hamuy01,elmhamdi03,leonard02a,leonard02b}. The \chem{56}{Ni} masses powering their light curve during the radioactive-decay phase are in the range $\sim$ 0.06-0.10\Msun \citep[e.g.][]{turatto90,sollerman02}. From a theoretical point of view, these features are explained in terms of core-collapse explosions with an energy of the order of 1 foe ($\equiv$10$^{51}$\,ergs), occurring in sufficiently massive progenitors (i.e.~stars with mass on main sequence larger than $\sim$ 8-10\Msun) that retain a substantial part (greater than $\sim$ 3-5 \Msun) of their hydrogen-rich envelope at the time of collapse \citep[e.g.][]{woosley86,hamuy03a,heger03,pumo09,PZ13}. 
Specifically the progenitors of SNe IIP are thought to be stars with initial (i.e.~at the zero age main sequence; ZAMS hereafter) mass up to $\sim$ 25-30\Msun\, that explode during the so-called red supergiant phase, even if the exact upper limit for the initial mass of the SNe IIP's progenitors is uncertain from both the theoretical and the observational point of view \citep*[e.g.][and references therein]{smartt09,we12,kochanek12}.\par

The discovery of SN 1997D and SN 1997D-like events (e.g.~SNe 1999br, 1999eu, 1999gn, 1994N, 2001dc, 2002gd, 2003Z, 2004eg, 2005cs, 2006ov, 2008bk and SN 2009md) has revealed the existence of a sub-group of SNe IIP with the following peculiar observational properties \citep*[e.g.][]{turatto98,benetti01,zampieri03,pasto04,pasto09,spiro14}: an under-luminous (at least a factor $\sim$ 5-10 lower than in normal SNe IIP) bolometric light curve at all epochs, a long lasting ($\gtrsim$ 100 days) plateau, and spectra with relatively narrow P-Cygni lines which are indicative of low expansion velocities  ($\lesssim$ 10$^3$ km s$^{-1}$ from the end of the plateau onwards and, in general, at least a factor $\sim$ 2-3 lower than in typical SN IIP explosions at any epoch) of the ejected material. 
These properties can be explained by Ni-poor (less than $\sim$ 10$^{-2}$\Msun), low-energy (of the order of tenths of foe) explosive events that seem to form the tail of a rather smooth distribution of SNe IIP. In fact, there is no evidence of a significant jump between the observed properties of these low-luminosity (LL) SNe IIP and the population of standard SNe IIP (see the early work of \citealt[][]{zampieri07} and \citealt[][]{spiro14} as well as the statistical studies of \citealt[][]{anderson14}, \citealt[][]{faran14} and \citealt[][]{sanders15}).\par

In contrast with other underluminous transients (e.g.~SN 2008S-like events, the archetypal SN impostor SN 1997bs and similar transients such as SNe 2002kg, 2003gm and 2007sv) whose own nature of actual sub-luminous SNe or non-terminal eruptions is still widely debated \citep*[e.g.][]{vandyk00,maund06,smith11,kochanek12b,AK15,tartaglia15,adams16}, LL SNe IIP seem to be genuine SN explosions with well-established general features. However, there are still uncertainties primarily linked to the real nature of their progenitors. 
Indeed theoretical arguments indicate the following three scenarios for the progenitors \citep[see][and references therein]{spiro14}: a) stars forming neon-oxygen degenerate cores (also known as super-asymptotic giant branch stars) sufficiently massive to evolve into a so-called electron-capture SN, b) low-energy explosions of red supergiant stars at the end of their quiescent evolution having initial masses below $\sim$ 15\Msun, and c) terminal explosions from more massive stars (i.e.~with initial masses $\gtrsim$ 20\Msun) where a non negligible fraction of the ejecta falls back onto the compact remnant (also known as fall-back SNe).\par

There are three approaches usually adopted to constrain the progenitors mass of observed LL SNe IIP: 1) the detection of the progenitor stars in pre-SN images  that allows a direct estimate of their masses, 2) the hydrodynamical modelling of the SN observables (i.e.~bolometric light curve, evolution of line velocities and continuum temperature at the photosphere), and 3) the modelling of the observed nebular spectra \citep[e.g.][]{jerkstrand12,jerkstrand14}. 
The first two methods, used more frequently, often produce discrepant results. Direct progenitor detections of LL SNe IIP provide ZAMS masses estimates in the range 8-15\Msun \citep*[see e.g.][]{maund05,li06,eldridge07,mattila08,smarttetal09,crockett11,fraser11,vandyk12,maund14a}, while the range of masses estimated from the hydrodynamical modelling is typically wider, including in some cases more massive progenitors \citep*[up to $\sim$ 20\Msun; see e.g.~][]{zampieri03,utrobin07,uc08}. 
For several events, however, the results from hydrodynamical modelling are in agreement with those obtained with the direct progenitor detection method \citep[see e.g.][]{zampieri07,pasto09, spiro14, takats14}. They best agree for the low-to-intermediate mass progenitors, pointing to low-energy explosions of red supergiant stars or super-asymptotic giant branch stars \citep[although the occurrence of electron-capture SNe from these stars is questioned; see e.g.][]{eldridge07}.
Nevertheless, both approaches used to constraining progenitor masses have their uncertainties and caveats. For the direct progenitor detection method, the main problems are: 1) uncertainties in the stellar evolutionary models (e.g.~treatment of mixing processes, rotation and mass-loss) used to infer the mass; 2) uncertainties in the extinction estimates \citep[][]{we12}, although \citet[][]{kochanek12} substantially reduce the relevance of this problem; and 3) possible selection effects because the method can be applied only to relatively close (within $\sim$ 30 Mpc) LL SNe IIP. For hydrodynamical modelling, the main caveats are a possible overestimate of the progenitor mass due to the one-dimensional approximation \citep[][]{uc09} and the poorly known pre-SN structure \citep[][]{dessart13}. 
All of this makes it difficult to progress in our knowledge on the LL SNe IIP's progenitors and, in particular, on the parameters describing the progenitor star at the time of the explosion, such as the progenitor radius at shock breakout, the ejected mass, and the explosion energy.\par

With the aim of clarifying the real nature of the LL SNe IIP's progenitors, we present the radiation-hydrodynamical modelling of three LL SNe IIP (namely, SNe 2003Z, 2008bk and 2009md). For these three well-observed SNe it is possible to derive reliable measurement of the physical parameters describing their progenitors at the time of the explosion. Together with the radiation-hydrodynamical modelling of other well-sampled LL SNe IIP presented in previous works \citep[SNe 2005cs, 2008in, 2009N, 2009ib and 2012A; see][]{spiro14,takats14,takats15,tomasella13}, we now have a sample of LL SNe IIP for which it is possible to carry out a comparative study based on the same modelling approach, enabling us to also identify possible systematic trends. A preliminary analysis of this type was carried out by \citet[][]{zampieri07} on a more limited sample of SNe IIP.\par

The plan of the paper is the following. We illustrate the radiation-hydrodynamical modelling procedure in Sect.~\ref{modelling} and shortly review the observational data in Sect.~\ref{sample}. In Sect.~\ref{results} we present and discuss our results, devoting Sect.~\ref{single_sne} to the three new objects and Sect.~\ref{systematics} to a comparative study of LL SNe IIP. A summary with final comments is presented in Sect.~\ref{summary}.\par

\section{Radiation-hydrodynamical modelling}
\label{modelling}

To perform the radiation-hydrodynamical modelling we use the same well-tested approach applied to other observed SNe (e.g.~2007od, 2009bw, 2009E, 2012aw, 2012ec and 2013ab; see \citealt{inserra11,inserra12}, \citealt{pasto12}; \citealt{dallora14}; \citealt{barbarino15}; and \citealt{bose15}, respectively). In this approach, the SN progenitor's physical properties at the explosion (namely the ejected mass $M_{ej}$, the progenitor radius at the explosion $R$ and the total explosion energy $E$) are constrained through the hydrodynamical modelling of all the main SN observables (i.e.~bolometric light curve, evolution of line velocities and the temperature at the photosphere), using a simultaneous $\chi^2$ fit of the observables against model calculations. 
It is well known that almost identical light curves can be obtained for more than one set of the parameters describing the SN progenitor's physical properties at the time of the explosion \citep[e.g.][]{arnett80,iwamoto98,nagy14}. This problem can, in turn, affect the search for possible correlations among the parameters, as correlations can be induced by covariance rather than by true physical effects \citep[][]{PP15}. For this reason, we try to reduce the ``degeneracy'' in the best-fitting model parameters by fitting simultaneously the evolution of the line velocity, the continuum temperature and the light curve.\par

Two codes are employed for computing the models. The first is a semi-analytic code that solves the energy balance equation for ejecta of constant density in homologous expansion \citep[see][for details]{zampieri03}. The second is a general-relativistic, radiation-hydrodynamics Lagrangian code that is
specifically designed to simulate the behavior of the main SN observables and the evolution of the physical properties of the ejected material at the time of the explosion from the breakout of the shock wave at the stellar surface up to the radioactive-decay phase (see \citealt*{pumo10} and \citealt{PZ11}, for details).
The distinctive features of this code are \citep[cf.~also][]{PZ13}: a) a fully general relativistic approach; b) an accurate treatment of radiative transfer coupled to hydrodynamics at all optical depth regimes; c) the coupling of the radiation moment equations with the equations of relativistic hydrodynamics during all the post-explosive evolution, adopting a fully implicit Lagrangian finite difference scheme; and d) a description of the evolution of the ejected material which takes into account both the gravitational effects of the compact remnant and the heating effects linked to the decays of the radioactive isotopes synthesized during the SN explosion.\par

The semi-analytic code is employed to carry out a preparatory analysis aimed at determining the parameter space describing the SN progenitor at the explosion. This guides the simulations performed with the general-relativistic, radiation-hydrodynamics code that are more realistic but time consuming.\par

We point out that the models are appropriate when the SN emission is dominated by the expanding ejecta with no significant contamination from interaction. In performing the $\chi^{2}$ fit, the observational data taken at the earliest phases (i.e.~within the first $\sim$ 20 days after explosion) are not included. This choice is made because the models could not accurately reproduce the early evolution of the main observables since the initial conditions used in the simulations are not able to precisely mimic the outermost high-velocity shell of the ejecta that forms after the shock breakout at the stellar surface \citep[see][for details]{PZ11}.\par

As for the comparison between the observations and the simulated SN observables, we remind the reader that the observed bolometric light curve is reconstructed from multi-color photometry and reddening measurements, whereas the photospheric velocity and temperature are estimated from the observed spectra \citep[see Sections 2.6, 3.4, and 5.1 of][for details on these procedures]{inserra11}. In particular, to estimate the photospheric velocity from the spectra, we use the minima of the profile of the Sc lines or, when they are not available, the Fe lines.\par

In the radiation-hydrodynamical modelling procedure the \chem{56}{Ni} mass initially present in the ejecta of the models is held fixed and its value is set so as to reproduce the observed bolometric luminosity during the radioactive decay phase. To this end, the initial \chem{56}{Ni} mass of the semi-analytic models is held fixed to that inferred from the observed late-time light curve. In the models computed with the general-relativistic, radiation-hydrodynamics code, the initial amount of \chem{56}{Ni} is in general larger, since the code accounts also for the material (including \chem{56}{Ni}) which falls back onto the compact remnant during the post-explosive evolution (see \citealt{PZ11} and the results of the modelling of SN 2009E in \citealt{pasto12}). 
However, in all the models presented here, the fall-back is negligible (a few hundredths of a solar mass) and, consequently, the initial \chem{56}{Ni} mass essentially coincides (within the errors) with that inferred from the observations. 
Other quantities held fixed in the radiation-hydrodynamical modelling procedure are the explosion epoch and the distance modulus. They are both used for determining the observed bolometric light curve and the evolution of the observed photospheric velocity and temperature as a function of phase.\par

\subsection{Uncertainties on the best-fitting model parameters}
\label{error}

The free model parameters of the fit are the ejected mass $M_{ej}$, the progenitor radius at the time of the explosion $R$ and the total explosion energy $E$. Although the evaluation of their uncertainties is not straightforward \citep[see also][]{zampieri03,uc09}, we estimate that the typical error due to the $\chi^{2}$ fitting procedure is $\sim$ 10-15\% for $M_{ej}$ and $R$ and $\sim$ 20-30\% for $E$. These errors are the 2-$\sigma$ confidence intervals for one parameter based on the $\chi^{2}$ distributions produced by the semi-analytical models. 
We used these models to determine the confidence intervals because it is needed a sufficiently high coverage of the parameter space, which is obtained through the calculation of thousands of models. The computation of such an extensive grid of models with the general-relativistic, radiation-hydrodynamics code is too expensive in terms of CPU time (e.g., running a single general-relativistic, radiation-hydrodynamics model takes up to $\sim$ 4-6 days).\par

The inferred uncertainties on the best-fitting model parameters do not include possible systematic errors related to the input physics (e.g.~opacity treatment, degree of \chem{56}{Ni} mixing and He/H ratio in the ejecta of the models) nor uncertainties on the assumptions made in evaluating the observational quantities (e.g.~the adopted reddening, explosion epoch and distance modulus). A discussion of the approximations on the input physics and the ensuing systematic errors can be found in \citet[][]{zampieri03}, while a detailed study of the effects of different opacity treatments on our radiation-hydrodynamical modelling will be presented in Pumo et al. (in prep.). 
Here, we recall that variations of the degree of \chem{56}{Ni} mixing and the He/H ratio mainly affect the plateau length of the models and not the simulated plateau luminosity or expansion velocity \citep[see e.g.][]{PZ13}. As a consequence, uncertainties related to these quantities should lead mainly to errors in the value of $M_{ej}$ as the plateau length depends mostly on it. Similar conclusions are also valid for uncertainties on quantities (e.g.~the adopted explosion epoch) that mainly affect the observed plateau length and, only to a secondary extent, the behaviour of the observed photospheric velocity and temperature.\par

Uncertainties related to the distance modulus or the adopted reddening basically produce a systematic variation in the brightness of the observed bolometric light curve at all phases and affect all three best-fitting model parameters. For example, the uncertainties on the reddening adopted for SN 2012aw modify the best-fitting model parameters obtained with our radiation-hydrodynamical modelling procedure by $\sim$ 15-30\% \citep[specifically $M_{ej}$, $R$ and $E$ vary up to $\sim$ 20\%, 30\% and 15\%, respectively; see][for details]{dallora14}. 
Somewhat larger changes in the best-fitting model parameters (specifically $M_{ej}$, $R$ and $E$ vary up to $\sim$ 25\%, 30\% and 35\%, respectively) are found after modifying the distance modulus of SN 2005cs from the value of 29.26 assumed by \citet[][]{pasto09} to the one of 29.46 adopted in \citet[][]{spiro14}. Although in these two test cases the variations of the best-fitting model parameters are significant, they do not have a dramatic impact on the overall results and the progenitor scenario.\par

\section{Sample of modelled underluminous IIP SNe}
\label{sample}

We model the well-studied LL SNe IIP 2003Z, 2008bk and 2009md. All the observational data used in the present work are taken from \citet[][and references therein]{spiro14}, where the observational features of these LL SNe IIP as well as a detailed description of the data reduction techniques have been extensively presented and discussed.\par

\begin{table}
   \centering
   \caption{Basic parameters (see text for details) for the sample of modelled LL SNe IIP.}
   \begin{tabular}{l ccc}
   \hline\hline
   SN                                      & 2003Z   & 2008bk   & 2009md \\
   \hline
   Adopted explosion epoch [JD]            & 52665   & 54550    & 55162  \\
   Adopted distance modulus                & 31.70   & 27.68    & 31.64  \\
   Estimated mass of \chem{56}{Ni} [\msun] & 0.005   & 0.007    & 0.004  \\
  \hline\hline
 \end{tabular}
 \label{tab_sample}
  All reported quantities are taken from Table 13 of \citet[][]{spiro14} that, in turn, used data from \citet[][]{mattila08} and \citet[][]{vandyk12} for SN 2008bk, and from \citet[][]{fraser11} for SN 2009md.
\end{table}

To be thorough, here (see Table \ref{tab_sample}) we recall the main assumptions made for evaluating the bolometric light curve and obtaining the behavior of the photospheric velocity and temperature as a function of the phase (i.e.~the adopted explosion epoch and distance modulus) as well as the amount of \chem{56}{Ni} estimated through a comparison with the late-time luminosity of SN 1987A. In Figures \ref{fig:03Z}, \ref{fig:08bk} and \ref{fig:09md} we also show the modelled observables (see the green squares) for SNe 2003Z, 2008bk and 2009md, respectively.\par

\section{Results and discussion}
\label{results}

\subsection{Individual objects}
\label{single_sne}

\begin{figure}
 \includegraphics[angle=-90,width=85mm]{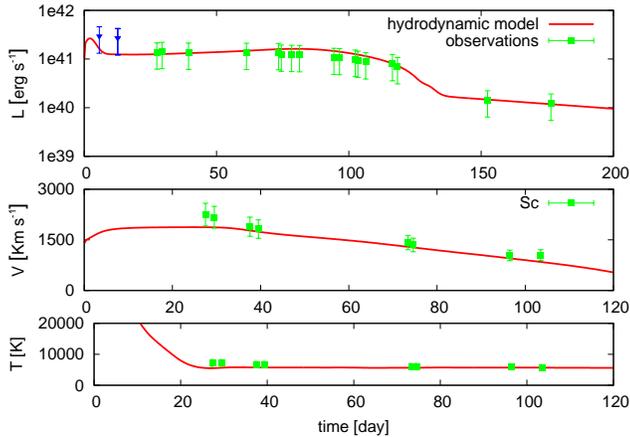} 
 \caption{Comparison of the evolution of the main observables of SN 2003Z with the best-fit model computed with the general-relativistic, radiation-hydrodynamics code. The best-fit model parameters are: total energy $0.16$ foe, radius at explosion $1.8 \times 10^{13}$ cm, and ejected mass $11.3$ \Msun. Top, middle, and bottom panels show the bolometric light curve, the photospheric velocity, and the photospheric temperature as a function of time. Blue triangles mark ``early'' observations not considered in the fitting procedure (see Sect.~\ref{modelling} for details).
 \label{fig:03Z}}
\end{figure}

\begin{figure}
 \includegraphics[angle=-90,width=85mm]{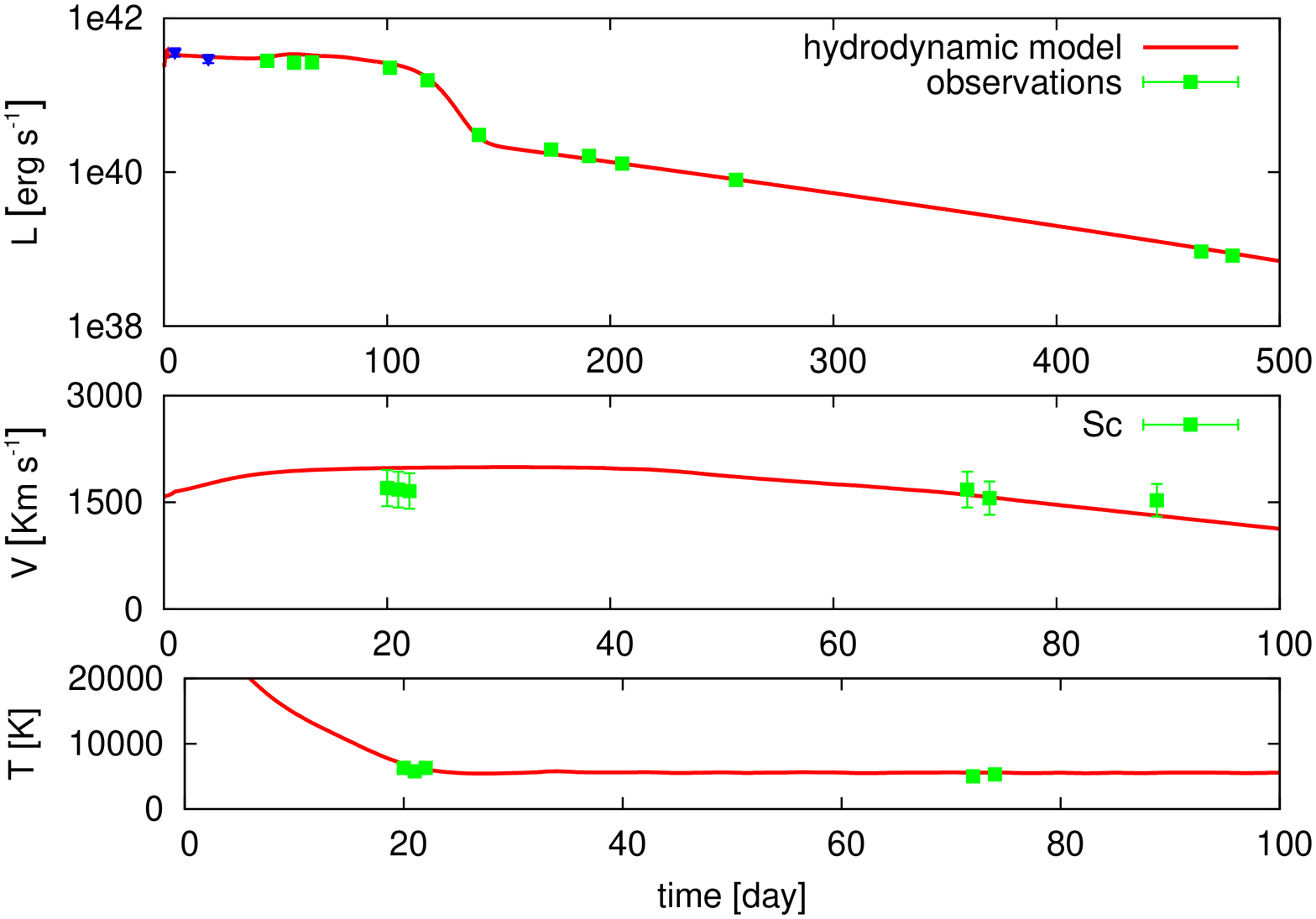} 
 \caption{Same as Fig.~\ref{fig:03Z}, but for SN 2008bk. The best-fit model parameters are: total energy $0.18$ foe, radius at explosion $3.5 \times 10^{13}$ cm, and ejected mass $10$ \Msun.
 \label{fig:08bk}}
\end{figure}

\begin{figure}
 \includegraphics[angle=-90,width=85mm]{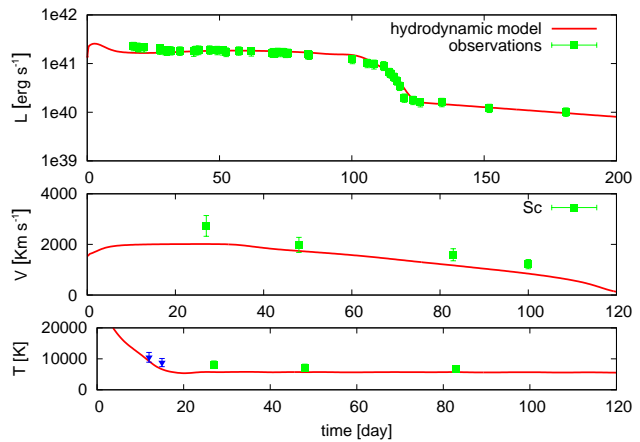} 
 \caption{Same as Fig.~\ref{fig:03Z}, but for SN 2009md. The best-fit model parameters are: total energy $0.17$ foe, radius at explosion $2 \times 10^{13}$ cm, and ejected mass $10$ \Msun.
 \label{fig:09md}}
\end{figure}

The best-fitting models for SNe 2003Z, 2008bk and 2009md are shown in Figures \ref{fig:03Z}, \ref{fig:08bk} and \ref{fig:09md}, respectively. The estimated uncertainties on the best-fitting model parameters are $\sim$ 10-15\% for $M_{ej}$ and $R$ and $\sim$ 20-30\% for $E$ (cf.~Section \ref{error}). To estimate the total stellar mass at the time of the explosion we consider a mass of $\sim$ 1.3-2\msun\, for the compact remnant \citep[e.g.][]{demorest10,sukhbold16}. To evaluate the ZAMS mass, we add to the inferred value of the total stellar mass at the time of the explosion an estimate of the mass lost during the pre-SN evolution based on the non-rotating stellar models reported in the recent literature \citep*[e.g.][]{heger00,hirschi04,pumo09,CL13}.\par

For SN 2003Z, the  $\chi^2$ fit procedure returns a best-fit model with a total (kinetic plus thermal) energy of 0.16 foe, a radius at the time of the explosion of 1.8 $\times$ 10$^{13}$ cm ($\sim$ 260\rsun) and an ejected mass of 11.3\msun. Adding the mass of the compact remnant to that of the ejected material, we obtain a total stellar mass at the time of the explosion of $\sim$ 12.6-13.3\msun. Such a mass and the other best-fit model parameters $R$ and $E$ are consistent with a low-energy explosion of a low-mass red supergiant star. The inferred radius could also indicate a yellow supergiant star as the progenitor of SN 2003Z, as it has been hypothesized for other SNe IIP \citep[e.g.~SNe 2004et and 2008cn;][]{li05,EliasRosa09} including some underluminous events \citep[e.g.~SN 2009N;][]{takats14}.
Assuming that $\sim$ 0.6-1.8\Msun\, are lost during the whole (i.e.~main sequence plus red/yellow supergiant phase) pre-SN evolution (see models with pre-SN mass close to $\sim$ 13\msun), the progenitor mass of SN 2003Z on the ZAMS is in the range $\sim$ 13.2-15.1\Msun. 
Unfortunately, there is no independent estimate of the SN 2003Z progenitor's initial mass, so it is not possible to make a direct comparison with our results. However, the value we infer for the progenitor mass of SN 2003Z on the ZAMS is comparable to the low progenitor masses found from observations of the progenitors of other LL SNe IIP (cf.~Sect.~\ref{intro}). Interestingly, with their hydrodynamical model, \citet[][]{utrobin07} derive a ZAMS mass of 14.4-17.4\Msun\, for the progenitor of SN 2003Z, consistent with our estimate, although they find a wider overall range. The values of all our other  derived parameters are also close to those of \citet[][]{utrobin07}, including the radius at explosion that led \citet[][]{utrobin07} to hypothesize a yellow supergiant star as the progenitor of SN 2003Z.\par 
                                                        
For SN 2008bk, the inferred best-fit model has a total energy of $0.18$ foe, a radius at explosion of $3.5 \times 10^{13}$ cm ($\sim$ 500\rsun) and an ejected mass of 10\msun. Adding the mass of the compact remnant, we obtain a total stellar mass at explosion of $\sim$ 11.3-12\msun. These values are fully consistent with the explosion of a low-mass red supergiant star, even if they may be also marginally consistent (within the errors) with an explosion of a super-asymptotic giant branch star with an initial mass close to the upper limit of the mass range typical of this class of stars, M$_{mas}$ \citep[see][and references therein]{pumo09}. 
Considering that the mass lost during the pre-SN evolution is $\sim$ 0.6-0.9\Msun\, for an exploding low-mass red supergiant star (see models with pre-SN mass close to $\sim$ 11.5\msun) or $\sim$ 0.1-0.3\Msun\, for a super-asymptotic giant branch star with an initial mass close to M$_{mas}$, the progenitor mass of SN 2008bk on the ZAMS is in the range $\sim$ 11.4-12.9\Msun\, fully in agreement with the estimate of 11.1 to 14.5\msun\, from the direct progenitor detection method by \citet[][]{maund14b}.\par


For SN 2009md, the best-fit model has a total energy of $0.17$ foe, a radius at explosion of $2 \times 10^{13}$ cm ($\sim$ 290\rsun) and an ejected mass of 10\msun. The estimated total stellar mass at the time of the explosion is $\sim$ 11.3-12\msun, consistent with the explosion of a low-mass red supergiant star. 
However, as in SN 2003Z, the  estimated radius at the time of the explosion could also suggest a yellow supergiant star as progenitor. Moreover, the best-fit model parameters may be also marginally consistent (within the errors) with the explosion of a super-asymptotic giant branch star with an initial mass close to M$_{mas}$. Using the same values of mass loss adopted for SN 2008bk, we find that the progenitor mass of SN 2009md on the ZAMS is in the range $\sim$ 11.4-12.9\Msun\, once again in agreement with the value (ranging from 7 to 15\msun) inferred from modelling the progenitor \citep[][]{fraser11}, although the identification of the real progenitor star of SN 2009md in the pre-SN images should be taken with caution \citep[][]{maund15}.\par

\subsection{Underluminous IIP SNe: a comparative analysis}
\label{systematics}

\begin{figure}
 \includegraphics[angle=-90,width=85mm]{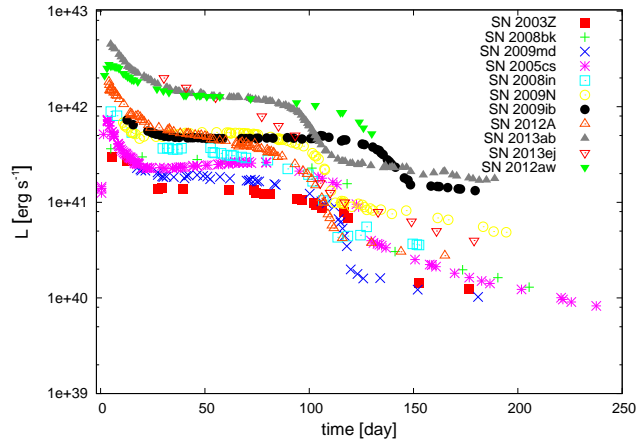}
 \caption{(Pseudo-)Bolometric luminosities during the first 250 days after the explosion for the SNe IIP reported in Table \ref{tab_systematics}. Data are taken from \citet[][]{spiro14} for SNe 2003Z, 2008bk and 2009md (cf.~Sect.~\ref{sample}) and from the papers reported in the last column of Table \ref{tab_systematics} for the remaining SNe.
 \label{fig:bol}}
\end{figure}

SNe 2005cs, 2008in, 2009N, 2009ib and 2012A \citep[whose radiation-hydrodynamical models were presented in previous works; see][]{spiro14,takats14,takats15,tomasella13} along with SNe 2003Z, 2008bk and 2009md (whose radiation-hydrodynamical models were presented in Sect.~\ref{single_sne}) form a sample of well-observed LL SNe IIP modelled in the same way. For them it has been thus possible to derive reliable and homogeneous estimates of the physical parameters describing the progenitors at the time of the explosion (see Table \ref{tab_systematics}). 
Note also that the sample is composed of both faint SNe IIP (namely, SNe 2003Z, 2005cs, 2008bk and 2009md) with bolometric luminosity at the plateau less than $\sim$~3~$\times 10^{41}$~erg~s$^{-1}$ and so-called ``intermediate-luminosity'' objects (namely, SNe 2008in, 2009N, 2009ib and 2012A) that are located between faint and standard SNe IIP (see Figure \ref{fig:bol}). 
All of this enables us to compare these LL SNe IIP, making possible the identification of potential systematic trends inside this sub-group of SNe IIP.\par

\begin{table*} 
   \centering
   \caption{Best-fitting model parameters and selected observational quantities.}
   \begin{tabular}{l ccccccc}
   \hline\hline
   \multicolumn{8}{c}{\normalsize LL SNe IIP}\\
   \multicolumn{8}{c}{Faint objects}\\
   SN     & $E$   & $M_{ej}$ & $R$    & $E/M_{ej}$  & \chem{56}{Ni}       & L$_{50}$               & Ref.               \\
          & [foe] & [\Msun]  & [cm]   & [foe/\Msun] & [\Msun]             & [erg sec$^{-1}$]       &                    \\
   2003Z  & 0.16  & 11.3     & 1.8e13 & 0.014       & 0.005  ($\pm$0.003) & 1.36e41 ($\pm$7.58e40) & this work          \\   
   2008bk & 0.18  & 10.0     & 3.5e13 & 0.018       & 0.007  ($\pm$0.001) & 2.76e41 ($\pm$3.45e40) & this work          \\
   2009md & 0.17  & 10.0     & 2.0e13 & 0.017       & 0.004  ($\pm$0.001) & 1.88e41 ($\pm$3.70e40) & this work          \\
   2005cs & 0.16  &  9.5     & 2.5e13 & 0.017       & 0.006  ($\pm$0.003) & 2.43e41 ($\pm$1.35e41) & \citet{spiro14}    \\
   \multicolumn{8}{c}{Intermediate-luminosity objects}\\  
   SN     & $E$   & $M_{ej}$ & $R$    & $E/M_{ej}$  & \chem{56}{Ni}       & L$_{50}$               & Ref.               \\
          & [foe] & [\Msun]  & [cm]   & [foe/\Msun] & [\Msun]             & [erg sec$^{-1}$]       &                    \\
   2008in & 0.49  & 13.0     & 1.5e13 & 0.038       & 0.012  ($\pm$0.005) & 3.71e41 ($\pm$1.36e41) & \citet{spiro14}    \\
   2009N  & 0.48  & 11.5     & 2.0e13 & 0.042       & 0.020  ($\pm$0.004) & 5.40e41 ($\pm$1.58e41) & \citet{takats14}   \\
   2009ib & 0.55  & 15.0     & 2.8e13 & 0.037       & 0.046  ($\pm$0.015) & 4.67e41 ($\pm$1.14e41) & \citet{takats15}   \\
   2012A  & 0.48  & 12.5     & 1.8e13 & 0.038       & 0.011  ($\pm$0.004) & 4.93e41 ($\pm$6.97e40) & \citet{tomasella13}\\
  \hline
  \multicolumn{8}{c}{\normalsize Standard-luminosity SNe IIP}\\
   SN     & $E$   & $M_{ej}$ & $R$    & $E/M_{ej}$  & \chem{56}{Ni}       & L$_{50}$               & Ref.               \\
          & [foe] & [\Msun]  & [cm]   & [foe/\Msun] & [\Msun]             & [erg sec$^{-1}$]       &                    \\
   2013ab & 0.35  & 7.0      & 4.2e13 & 0.050       & 0.06   ($\pm$0.003) & 1.30e42 ($\pm$2.60e41) & \citet{bose15}     \\
   2013ej & 0.70  & 10.6     & 4.2e13 & 0.066       & 0.02   ($\pm$0.01)  & 1.04e42 ($\pm$2.08e41) & \citet{huang15}    \\
   2012aw & 1.50  & 19.6     & 3.0e13 & 0.079       & 0.06   ($\pm$0.013) & 1.29e42 ($\pm$2.58e41) & \citet{dallora14}  \\
  \hline
  \hline
  \end{tabular}
 \label{tab_systematics}
 
 The quantities shown are (from left to right): the SN name, the best-fitting model parameters $E$, $M_{ej}$ and $R$ (cf.~Sect.~\ref{modelling}), the ratio of $E$ to $M_{ej}$, the fixed \chem{56}{Ni} mass inferred from the observations (see Sect.~\ref{modelling}), the plateau (pseudo-)bolometric luminosity (measured at 50 days after explosion and derived by interpolating the data shown in Figure \ref{fig:bol}), and the reference to the paper where the radiation-hydrodynamical models are presented in detail. Estimated uncertainties on the values inferred from the observations are in brackets. Top and bottom panels refer respectively to the sample of LL SNe IIP and to three SNe IIP of standard luminosity (namely, SNe 2013ab, 2013ej and 2012aw) modelled in the same way (i.e.~using the approach described in Sect.~\ref{modelling}), which are reported for the sake of comparison.
\end{table*}

As it can be seen in the top panel of Table \ref{tab_systematics}, all the model parameters (with the only exception of $R$) as well as the ratio $E$/$M_{ej}$ and the \chem{56}{Ni} mass of the faint SNe IIP are systematically lower than those for the intermediate-luminosity objects, showing that the progenitors of the faint SNe IIP are slightly less massive and experience less energetic explosions than the progenitors of the intermediate-luminosity objects.\par

Moreover, our radiation-hydrodynamical models suggests that the present sample of faint SNe IIP and intermediate-luminosity objects originate from stars of low-to-intermediate mass, in agreement with the results found for some of them by modelling their progenitors. In particular we find that the best-fit model parameters of all the modelled LL SNe IIP are consistent with low-energy explosions of red (or yellow) supergiant stars and, for some faint objects \citep[SNe 2009md and 2008bk as well as SN 2005cs, whose model is presented in][]{spiro14}, they can also be consistent with explosions of massive, super-asymptotic giant branch stars as electron-capture SNe.\par

\begin{figure}
 \includegraphics[angle=-90,width=85mm]{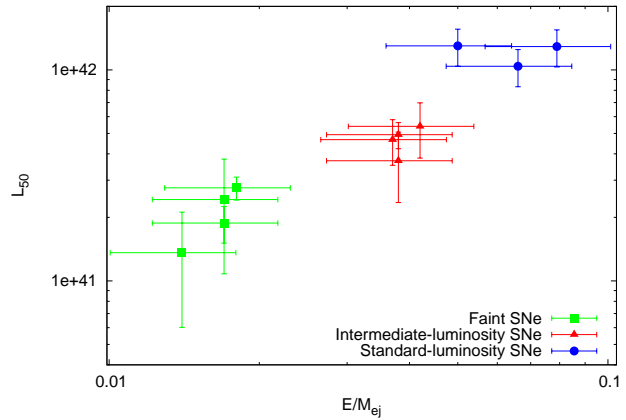}
 \caption{Correlation between the plateau (pseudo-)bolometric luminosity and the ratio $E/M_{ej}$ for the SNe IIP reported in Table \ref{tab_systematics}. The plateau luminosity is measured at 50 days after the explosion and derived by interpolating the data shown in Figure \ref{fig:bol}. Its errorbars are coincident with the values reported in Table \ref{tab_systematics} which were inferred from the observations. The errorbars on the $E/M_{ej}$ ratios are estimated by propagating the uncertainties on $E$ and $M_{ej}$, adopting a value of 30\% for the relative error of $E$ and 15\% for that of $M_{ej}$ (cf.~Sections \ref{error} and \ref{single_sne}).
\label{fig:correl1}}
\end{figure}

\begin{figure}
 \includegraphics[angle=-90,width=85mm]{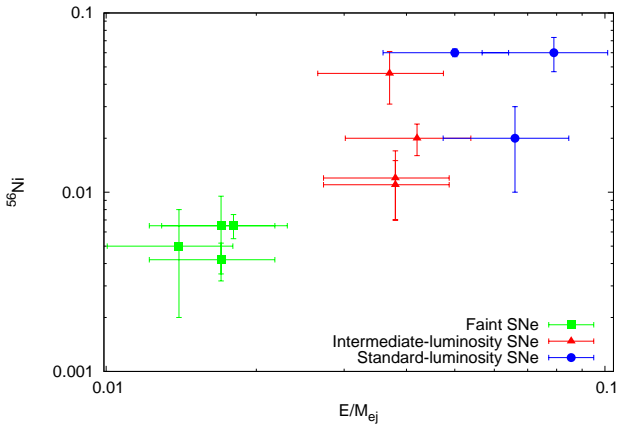}
 \caption{Same as Fig.~\ref{fig:correl1}, but for the correlation between the \chem{56}{Ni} mass and the ratio $E/M_{ej}$. The errorbars on the \chem{56}{Ni} masses are the value reported in Table \ref{tab_systematics}, inferred from the observations. The errorbars on the $E/M_{ej}$ ratios are evaluated as described in the caption of Fig.~\ref{fig:correl1}.
 \label{fig:correl2}}
\end{figure}

The data reported in Table \ref{tab_systematics} and Figures \ref{fig:bol} to \ref{fig:correl2} also confirm that LL SNe IIP form the underluminous tail of the family of SNe IIP \citep[see also][]{hamuy03b,pasto04,zampieri07,spiro14,anderson14,faran14,sanders15}. With the warning that our sample could be still too small to draw final conclusions, the main parameter ``guiding'' the distribution seems to be the ratio of $E$ to $M_{ej}$, not just the explosion energy $E$. Indeed, Figures \ref{fig:correl1} and \ref{fig:correl2} reveal a relationship between the observed quantities such as the plateau luminosity and the \chem{56}{Ni} mass and the ratio $E/M_{ej}$, which monotonously increases from $\sim$ 0.015-0.02 to $\sim$ 0.04 up to values $\gtrsim$ 0.05, when passing from faint to intermediate-luminosity up to ``standard-luminosity'' events, respectively.\par

It is also clear that the data, in particular the correlation of \chem{56}{Ni} with $E/M_{ej}$ in Figure \ref{fig:correl2}, show significant scatter. Indeed, although the explosion energy as well as the other model parameters and the \chem{56}{Ni} mass tend to increase when moving from LL SNe IIP to SNe IIP of standard luminosity, there are several exceptions (see Table \ref{tab_systematics}). 
For example, SN 2013ab, a SN IIP with standard luminosity, is characterized by an explosion energy of 0.35 foe, significantly lower than the explosion energies of standard SNe IIP and between those of faint SNe and the ones of intermediate-luminosity objects. Also the ejected mass is small and has the lowest value of the sample of SNe IIP reported in Table \ref{tab_systematics}. On the other hand, the ratio $E/M_{ej}$, the radius at the explosion and the \chem{56}{Ni} mass are similar to those of standard SNe IIP, explaining the normal luminosity of this object. 
Another exception is SN 2009ib \citep[see also][]{takats15}, an intermediate-luminosity object with \chem{56}{Ni} mass closer to that of normal SNe IIP, explosion energy slightly higher than the one of the other intermediate-luminosity objects and ejected mass among the highest of the sample of SNe IIP reported in Table \ref{tab_systematics}. However the ratio $E/M_{ej}$ is very close to that of the other intermediate-luminosity objects, explaining the intermediate luminosity of this SN. Other minor outliers are: SN 2008bk with its relatively large radius at explosion and SN 2013ej with a value of \chem{56}{Ni} mass closer to that of intermediate-luminosity objects and an ejected mass similar to that of faint SNe.\par


\section{Summary and  further comments}
\label{summary}

In order to improve our knowledge of the real nature of the progenitors of LL SNe IIP, we made radiation-hydrodynamical models of the well-studied SNe 2003Z, 2008bk and 2009md members of this sub-group of explosive events. We used the same well-tested approach applied to several other observed SNe (e.g.~2007od, 2009bw, 2009E, 2009N, 2009ib, 2012A, 2012aw, 2012ec and 2013ab; see \citealt{inserra11,inserra12}, \citealt{pasto12}; \citealt{takats14,takats15}; \citealt{tomasella13}; \citealt{dallora14}; \citealt{barbarino15}; and \citealt{bose15}, respectively). In this approach, the SN progenitor's physical properties at explosion (namely the ejected mass $M_{ej}$, the progenitor radius at the time of the explosion $R$ and the total explosion energy $E$) are constrained by modelling the bolometric light curve, the evolution of line velocities and the temperature of the photosphere, and performing a simultaneous $\chi^2$ fit of the model calculations to these observables.\par

The inferred parameters describing the SN progenitors and their ejecta for SNe 2003Z, 2008bk and 2009md are fully consistent with low-energy explosions of red supergiant stars with relatively low mass, although the value of $R$ could also suggest yellow supergiant stars as the progenitors of SNe 2003Z and 2009md. The best-fitting model parameters inferred for SNe 2008bk and 2009md may also be consistent with the explosion of super-asymptotic giant branch stars with initial masses close to the upper limit of the mass range typical of this class of stars.\par

Assuming a mass of $\sim$ 1.3-2 \msun\, for the compact remnant and a ``standard'' (i.e.~not enhanced by rotation) mass loss during the pre-SN evolution, we estimate that the progenitor masses on the ZAMS are in the range  $\sim 13.2$-$15.1$\Msun\, for SN 2003Z and in the range $\sim 11.4$-$12.9$\Msun\, for SNe 2008bk and 2009md. The latter two estimates agree with those based on direct observation of the progenitors.\par

Since these results were obtained in the same way as those for SNe 2005cs, 2008in, 2009N, 2009ib and 2012A, we can conduct a comparative study on this sub-group of SNe IIP. The main findings of this comparative analysis can be summarized as follows:
\begin{itemize}
 \item the progenitors of faint SNe IIP are slightly less massive and experience less energetic explosions than the progenitors of the intermediate-luminosity objects, even though both faint SNe IIP and intermediate-luminosity objects originate from low-energy explosions of red (or yellow) supergiant stars of low-to-intermediate mass;
 \item some faint SNe IIP may also be explained as electron-capture SNe involving massive super-asymptotic giant branch stars, although the existence of such explosive events is still not completely proven;
 \item LL SNe IIP form the underluminous tail of the family of SNe IIP where the main parameter ``guiding'' the distribution seems to be the ratio $E/M_{ej}$. 
\end{itemize}

Admittedly, our sample is still too small to draw final conclusions. For this reason, other studies based on a larger sample of LL SNe IIP, including also more extreme events as SNe 1999br-like, are needed to confirm these findings. Moreover it should be useful to further check the results performing the radiation-hydrodynamical modelling in a ``self-consistent'' way, that is to say using numerical calculations which include the SN explosion and the explosive nucleosynthesis, and that start from pre-SN models evaluated by means of stellar evolution codes \citep[see][for further details]{PZ11,PZ12}.\par

Although none of the LL SN IIP in our sample appears to be well modelled with massive ejecta and/or explained in terms of a fall-back SN, at present we cannot rule out that a minor fraction of their progenitors may be more massive than $\sim$ 15\msun\, and/or undergo significant fall-back. While a larger sample of LL SNe IIP is necessary to draw any firm conclusion, recent one-dimensional hydrodynamical simulations of neutrino-driven SNe \citep[][]{ugliano12,ertl16,sukhbold16} indicate that there is no monotonic progenitor mass dependence of the properties of core-collapse SNe. More specifically, for certain progenitor structures at explosion, not even particularly massive stars (15\msun $\lesssim$ initial mass $\lesssim$ 40\msun) could lead to direct collapse and the formation of a black hole \citep[see][]{OO13,PT15}. 
Thus, it is possible that extreme and comparatively rarer events with very low ejected \chem{56}{Ni} and explosion energy may hide among LL SNe IIP perhaps with more somewhat different observational properties and be explained in terms of almost failed explosions of not-particularly-massive stars undergoing significant fall-back, as suggested earlier by \citet[][]{zampieri03}. 
For example, in the simulations of \citet[][]{ugliano12} and \citet[][]{sukhbold16}, fall-back SNe occur in only a few cases for progenitor stars with initial masses in the range $\sim$ 25-40\msun\, and, in these cases, the explosion properties (i.e.~explosion energy, ejected mass, and amount of \chem{56}{Ni}) seem to be qualitatively similar to those of observed LL SNe IIP, albeit somewhat more extreme. However, as observed by the same authors \citep[see e.g.][]{ugliano12}, the results of such simulations must be used with caution for a direct comparison with the observed properties of LL SNe IIP, because 1) they are based on sets of progenitor models which, even for similar initial masses, exhibit large structural variations that may not be completely realistic, and 2) the simulations do not consider multidimensional effects that can play a critical role in the core-collapse SN mechanism \citep[see also][]{ertl16}. 
So additional multidimensional hydrodynamical studies focused on the progenitor-explosion and progenitor-remnant connections should be performed and compared to a sufficiently extended sample of LL SNe IIP before drawing any final conclusion on the possible occurrence of fall-back in some LL SNe IIP. 
Observationally, a statistically sound test on the existence of progenitors having significant fall-back will come by searching for failed SNe \citep*[see e.g.~][]{kochanek08,gerke15} and from larger numbers of direct progenitor detections. If a stringent mass progenitor upper limit of $\sim$ 15\msun\, will be established, then this would severely limit the occurrence of such a process. In such a case, either successful and completely failed explosions form two ``final states'' separated by a sharp transition, the outcome of which depending on fine details of the internal structure of their progenitors, or almost failed explosions are not seen because they are typically too weak to be detected in present surveys.

\section*{Acknowledgments}
M.L.P.~acknowledges the financial support from INAF-OAPA and CSFNSM. AP, SB, EC, MT are partially supported by the PRIN-INAF 2014 project ``Transient Universe: unveiling new types of stellar explosions with PESSTO''. We thank an anonymous referee for his/her valuable comments and suggestions that improved our manuscript.

\bibliographystyle{aa}

\label{lastpage}

\end{document}